\documentclass[preprint,prl,aps]{revtex4}
\usepackage{amsmath}
\usepackage{amssymb}

\newcommand{\bse}{\begin{subequations}}
\newcommand{\ese}{\end{subequations}}

 \newcommand{\be}{\begin{equation}}
 \newcommand{\ee}{\end{equation}}
 \newcommand{\ben}{\begin{eqnarray}}
 \newcommand{\een}{\end{eqnarray}}
 \newcommand{\ii}{\'{\i}}
 
 \newcommand{\nd}{\noindent}

 \newcommand{\tr}{{\mathrm{Tr}}}
 
 \usepackage{amssymb}

\begin{document}

 \title{Equivalence of the four versions of Tsallis' statistics}
 \author{G. L. Ferri$^1$, S. Mart\'{\i}nez$^2$, and A.
 Plastino$^2$}

 \address{ $^1$ Facultad de Ciencias Exactas,
 Universidad Nacional de La Pampa, Argentina
 \\$^2$  La Plata Physics Institute (IFLP) \\
 National University La Plata and Argentine National
 Research
 Council (CONICET) \\  C.C. 727 , 1900 La Plata,
 Argentina }


 \begin{abstract}
 In spite of its undeniable success, there are still open questions
regarding Tsallis' non-extensive statistical formalism, whose
founding stone was laid in 1988 in JSTAT. Some of them are
concerned with the so-called {\it normalization} problem of just
how to evaluate expectation values. The Jaynes' MaxEnt approach
for deriving statistical mechanics is based on the adoption of (1)
a specific entropic functional form $S$ and (2) physically
appropriate constraints. The literature on non-extensive
thermostatistics has considered, in its historical evolution, four
possible choices for the evaluation of expectation values: (i)
1988 Tsallis-original (TO), (ii) Curado-Tsallis (CT), (iii)
Tsallis-Mendes-Plastino (TMP), and (iv) the same as (iii), but
using centered operators as constraints (OLM). The 1988 was
promptly abandoned and replaced, mostly with versions ii) and
iii). We will here (a) show that the 1988 is as good as any of the
others, (b) demonstrate that the four cases can be easily derived
from just {\it one} (any) of them, i.e., the probability
distribution function in each of these four instances may be
evaluated with a unique formula, and (c) numerically analyze some
consequences that emerge from these four choices.

 \vspace{0.2 cm}



 \vspace{1 cm}
 \end{abstract}
  \maketitle



 \newpage

 \section{I. Introduction}


\nd James Bernoulli published in the {\em Ars Conjectandi} the
first formal attempt to deal with probabilities in 1713 and
Laplace further formalized things in his {\em Th\'eorie analytique
des Probabilit\'es} of 1820. As we know, in the intervening years
Probability Theory (PT) has grown into a rich, powerful,  and
extremely useful branch of Mathematics. Contemporary Physics
heavily relies on PT for a large part of its basic structure,
Statistical Mechanics \cite{patria,reif,sakurai}, of course, being
a most conspicuous example. One of PT  basic definitions is that
of the mean value of an observable $\mathcal{A}$ (a measurable
quantity), which in its original form it is taken as a {\em lineal
combination} of the different probabilities that $\mathcal{A}$
adopts for the different accessible states (we denote them by the
index $i$). Let $A$ stand for the linear operator or dynamical
variable associated with $\mathcal{A}$. Then

\be \left<A\right>=\sum_i p_i A_i. \ee It is well known that, in
some specific cases, it  becomes necessary to use
 ``weighted" mean values, of the form
\be \left<A\right>=\sum_i f(p_i)\, A_i,  \ee with $f$ an
analytical function of the $p_i$. This happens, for instance, when
  there is a  set of states characterized by a distribution  with a
  recognizable maximum and a large tail that contains low but finite probabilities.
One faces then the need of making  a  pragmatical (usually of
experimental origin) decision regarding $f$.

\subsection{Nonextensive thermostatistics (NET)}

 \nd   Tsallis' nonextensive thermostatistics
(NET) is nowadays  considered  by many authors
 to be  a new
 paradigm for Statistical Mechanics (SM)
 ~\cite{tsallis,Brasil}. It encompasses
  a whole
 {\em family} of SM-formulations, all of them based upon
 Tsallis'
 nonextensive information measure \cite{tsallis,Brasil}

\be S_{q}=k_B\frac{1-\sum_{i=1}^{W}p_{i}^{q}}{q-1},
\label{tsallis}\ee where $k_{B}$ stands for Boltzmann's constant
and $\{p_{n}\}$ is a set of normalized probabilities. The real
parameter $q$ is called the index of nonextensivity. The
literature on Tsallis' thermostatistics considers three possible
choices for the evaluation of mean- or expectation-values (EVs)
within the nonextensive scenario. Since EVs are always regarded as
constraints in the associated MaxEnt approach \cite{katz}, three
types of $q-$statistics will thereby ensue. Consider the physical
quantity $O$ that, in the microstate $n$ $(n=1,2,...,W),$ adopts
the value $o_{n}$. Let $p_{n}$ stands for the microscopic
probability for such microstate. The  $O-$EV is evaluated
according to three distinct recipes, that we illustrate below with
reference to Gibbs'  canonical ensemble, where it is assumed that
the a priori physical information is given by the mean value of a
Hamiltonian ($O\equiv H$) of spectrum $\{o_i \equiv \epsilon_i\},$
and there are two Lagrange multipliers, one associated to
normalization ($\alpha$) and the other to the mean energy (inverse
temperature $\beta$).


i) First choice: Tsallis (1988) \cite{t0} \ \be \label{c1} \langle
O\rangle = \sum_{i=1}^{W}p_{i}o_{i}.\ee The associated,  Tsallis
canonical distribution was (see the pertinent demonstration below
with reference to Eq. (\ref{pii}))
 \ben \label{13} p_i&=&[(1-q)(\alpha+\beta \epsilon_i)/q]^{1/(q-1)}. \een
The awkward location of the multiplier $\alpha$ makes $p_i$ hard
to evaluate, save for very special cases like that of the harmonic
oscillator \cite{t0}. This motivated people to look for better
alternatives (see below). Other normalization choices thus emerged
that typically  employ ``weighted" mean values and avoid the
explicit presence of the multiplier $\alpha$:

ii) Second choice: Curado - Tsallis (1991) \cite{CT} \be
\label{c2} \langle O\rangle = \sum_{i=1}^{W}p_{i}^{q}o_{i},\ee
with a canonical distribution given by  \ben \label{17} p_i&=&
(Z_q)^{-1}
  \,\left[1-(1-q)\,\beta\, \epsilon_i \right]^{1/(1-q)} \cr
   Z_q&=& \sum_i\,\left[1-(1-q)\,\beta\, \epsilon_i \right]^{1/(1-q)}. \een

iii) Third choice: Tsallis - Mendes - Plastino (1998) \cite{TMP}
\be \label{c3} \langle O\rangle =
\frac{\sum_{i=1}^{W}p_{i}^{q}\,o_{i}}{\sum_{i=1}^{W}p_{i}^{q}}.
\ee \nd We need to define here a normalization factor
(\textsc{not} a Lagrange multiplier)

\be {\cal X}_q=\sum_i p_i^q \label{Xq}. \ee The pertinent
canonical distribution, that is {\it self-referential} because the
$p_i$ depends upon ${\cal X}_q$. Since it reads \cite{TMP} \ben
\label{105} p_i &=&
  Z_q^{-1}\,[1\,-\, \frac{(1-q)\beta }{{\cal
X}_q}
  \,(\varepsilon_i\,-\,U_q)]^{1/(1-q)} \cr
   {\cal X}_q&=& (Z_q)^{1-q}\,\,\,\,({\rm caracteristic\,\,identity})
   \cr     Z_q&=&\sum_i\,
   \left[1\,-\, \frac{(1-q)\beta }{{\cal
X}_q}\,(\varepsilon_i\,-\,U_q)\right]^{1/(1-q)},\een \nd this
means that even if one knows $U_q$ (as a real number) and the
spectrum (we call this the canonical practitioner's input), this
is not enough to obtain straightforwardly the probability
distribution. One still needs to repeatedly iterate (\ref{105})
until numerical self-consistency is achieved, which can be
difficult in practice. In order to overcome such a problem still
another choice can be made, the so-called OLM one
\cite{OLM,Rlenzi,FPS}, which avoids the self-reference problem.
The corresponding probability distribution arises from solving
anyone of two different variational problems \cite{sandra04},
namely, those arising after
\begin{enumerate}
\item
 replacing
$S_q$ by R\'enyi's information measure $S^{R}$ \be S^{R}=\frac{\ln
\sum_{i=1}^{W}p_{i}^{q}}{1-q}, \label{renyi} \ee and keeping the
TMP constraints. We call this procedure the TMP-R one.
\item Maintain
 Tsallis' entropy, but replace the TMP constraints by ``centered" mean
values. We call this procedure the OLM-one. \end{enumerate} \nd In
both cases one finds \cite{foot}

 \ben \label{Zrenyi} p_i &=&
  Z_q^{-1}\,[1\,-\, (1-q)\beta
  \,(\varepsilon_i\,-\,U_q)]^{1/(1-q)}
   \cr  {\cal X}_q&=& Z_q^{(1-q)}\,\,\,\,({\rm caracteristic\,\,identity})  \cr
Z_q&=&\sum_i\,
   \left[1\,-\, (1-q)\beta \,(\varepsilon_i\,-\,U_q)\right]^{1/(1-q)},\een
and it should be noted that, with access to the canonical
practitioner's input, no numerical self-consistency treatment is
now required.
\subsection{Reasons for using weighted mean values}

\nd In the first stage of NET development,
 its pioneer practitioners made the
pragmatic choice of using ``weighted" mean values, of rather
unfamiliar appearance for many physicists. Why? The reasons were
of theoretical origin. It was at the time believed that, using the
familiar linear, unbiased mean values, one was unable to get rid
of the Lagrange multiplier associated to probability-normalization
(see below), usually denoted with the letter $\alpha$ in the
Literature. Since the Tsallis formalism yields, in the limit
$q\rightarrow 1$, the orthodox Jaynes-Shannon
 treatment, the natural  choice was to construct weighted EVs using the
  index  $q$

\be \left<A\right>_q=\sum_i p_i^q A_i, \ee the so-called
Curado-Tsallis unbiased mean values (MV) \cite{CT}. As shown by
Plastino and Plastino \cite{pp1}, employing the Curado-Tsallis
(CT)  mean values allowed one to obtain an {\it analytical}
expression for the partition function out of the concomitant
MaxEnt process \cite{pp1}. This EV choice leads to a non extensive
formalism endowed with interesting
 features: i) the above mentioned property of its partition
 function $Z$, ii)
a  numerical treatment that is relatively simple, and iii) proper
results in the limit $q\rightarrow 1$.  It has, unfortunately, the
 drawback of exhibiting  un-normalized mean values, i.e.,
$\left<\left<1\right>\right>_q\neq 1$. The latter problem was
circumvented in the subsequent work of Tsallis-Mendes-Plastino
(TMP) \cite{TMP}, that ``normalized" the CT treatment by employing
mean values of the form

\be \left<A\right>_q=\sum_i \frac{p_i^q}{{\cal X}_q} A_i. \ee
 Today, this is the orthodox normalization
procedure for NET practitioners. However, the concomitant
treatment is not at all simple. Numerical  complications often
ensue, which has encouraged the development of different,
alternative  lines of NET research. See, for instance (a by no
means exhaustive list),
\cite{abe,abe1,almeyda,vives,disisto,toral,yamano} and references
therein.

\subsection{Present goals} \nd In this work we revisit the MaxEnt
approach
 with unbiased mean values as prior knowledge, and show that {\em it is possible} to obtain
 a probability distribution that  i) does not explicitly contain the obnoxious $\alpha$ factor,
ii) allows a seemingly  analytical expression for the partition
function out of the concomitant MaxEnt process (although it
remains self-referential), iii) it is normalized, and iv) leads to
the proper result in the $q\rightarrow 1$ limit. It does exhibit
some numerical drawbacks, similar to those faced by the TMP
treatment, but without the  theoretical limitations that affect
both the CT and the TMP formalisms. Since the treatment here
advanced allows one to avoid weighted expectation values, it might
be preferable to both the CT and and TMP ones on
 Ockham's razor-grounds.
\vskip 2mm
 \nd Moreover, we will  show that  all
four choices described above, namely, (\ref{c1}), (\ref{c2}),
(\ref{c3}), and OLM, are {\it equivalent} in the sense that
 the four cases can be easily derived from just {\it one}
(any) of them, i.e., the probability distribution function in each
of these four instances may be evaluated with a {\it unique}
formula, that (in the  canonical instance) reads, for a given
$q-$value \bse \label{igualidad} \ben p_{i}&=&\frac{\left[
1-\left( 1-q^{\ast }\right) \beta ^{\ast }\varepsilon _{i}\right]
^{1/(1-q^{\ast })}}{\sum_{i}\left[ 1-\left( 1-q^{\ast }\right)
\beta ^{\ast }\varepsilon _{i}\right] ^{1/(1-q^{\ast })}}\cr
&\equiv& \frac{\exp _{q}\left( -\beta ^{\ast }\varepsilon
_{i}\right) }{\sum_{i}\exp _{q}\left( -\beta ^{\ast }\varepsilon
_{i}\right) };\,\,{\rm for}\,\,\beta ^{\ast }\varepsilon
_{i}<\frac{1}{1-q^{\ast }}, \een \ben p_{i}&=&0; \,\,\,\,{\rm
if}\,\,\,\, \beta ^{\ast }\varepsilon _{i} \ge
 \frac{1}{1-q^{\ast }},\een \ese where $\beta ^{\ast }$ is an
 appropriate
pseudo-inverse temperature ($=1/k_{B}T^{\ast }$ ) which depends on
$\beta, $ ${\rm exp}_q$ is the well-known $q-$generalized
exponential \cite{tsallis} \be {\rm exp}_q(x) \equiv
[1+(1-q)x]^{1/(1-q)}, \label{expq} \ee  and $q^{\ast }$ is an
effective nonextensivity index, suitably related to  $q$ (see
Table 1). The extension of (\ref{igualidad}) to the general case
of $M$ observables ${\cal O}_j,\,\,(j=1,\ldots,M)$, that involves
$M$ Lagrange multipliers $\lambda_j,\,\,(j=1,\ldots,M)$, is
straightforward.

 \section{II. RESCUING TSALLIS 1988 VERSION}
 \subsection{Classical canonical treatment}
\nd  For didactic purposes we will restrict ourselves first to the
canonical ensemble and the classical formulation. The forthcoming
section will treat things in quantal fashion for general
ensembles. The Boltzmann constant will be taken as unity in this
Section. We  write the internal energy (the mean value of the
Hamiltonian) as the linear probabilities combination

\be U_q=\langle H\rangle=\sum_{i=1}^W p_i H_i, \ee where the sum
is taken over the $W$ micro-states of the system. The MaxEnt
procedure entails maximizing Tsallis information measure subject
to the constraints given by the internal energy and normalization
of the probabilities

\be \sum_i p_i=1. \label{norm} \ee The quantity to be maximized is

\be L_T= S_q-\alpha\sum_i p_i-\beta U_q, \ee i.e., \be
\frac{\partial L_T}{\partial
p_i}=\frac{q}{1-q}p_i^{q-1}-\alpha-\beta H_i=0, \label{1pi} \ee
wherefrom we easily find

\be p_i=\left[\frac{(1-q)}{q}(\alpha+\beta H_i)\right]^{1/(q-1)}.
\label{pii} \ee

\nd This is the result obtained and used in the very first paper
on the Tsallis formalism \cite{t0}, and subsequently {\it
abandoned} i) because of the presence of the $\alpha$ term in
(\ref{pii}) and ii) due to the fact that it was seemingly
impossible to obtain a proper partition function. The step that
transforms the above treatment into a useful one is performed via
a simple trick. Note that we can obtain $\alpha$ by first
multiplying Eq. (\ref{1pi}) by $p_i$ and then summing over $i$,
which yields

\be \alpha=\frac{q}{1-q}{\cal X}_q-\beta U_q. \label{alpha} \ee
Once we have $\alpha$, the problem is solved. If we place $\alpha$
 into Eq. (\ref{pii}) we obtain \be
p_i=\left[{\cal X}_q-\frac{q-1}{q}\beta \delta
H_i\right]^{1/(q-1)}, \label{2pi} \ee where $\delta H_i=H_i-U_q$.
To explicitly obtain the partition function is now a  simple task.
 Using Eq. (\ref{norm}) one immediately finds

\be 1=\frac{\sum_i\left[1-\frac{q-1}{q}\frac{\beta}{{\cal X}_q}
\delta H_i\right]^{1/(q-1)}}{{\cal X}_q^{1/(1-q)}}, \ee so that,
if we agree to call

\be \bar{Z}_q=\sum_i\left[1-\frac{q-1}{q}\frac{\beta}{{\cal X}_q}
\delta H_i\right]^{1/(q-1)}, \ee we obtain the well-known, typical
 TMP result \cite{TMP}
\be {\cal X}_q=\bar{Z}_q^{1-q}, \label{XZrel} \ee and

\be p_i=\frac{\left[1-(q-1)\frac{\beta}{q{\cal X}_q} \delta
H_i\right]^{1/(q-1)}}{\bar{Z}_q}. \ee

\nd It is clear that the need of having a real probability forces
one to retain only the positive values of the expression between
brackets (cut-off condition). Notice that the result (\ref{XZrel})
resembles the TMP one and that the  presence of the quantity
${\cal X}_q$ makes the probability distribution a self referential
one, although  the proper limit when $q\rightarrow 1$ is indeed
obtained \be \left. p_i\right|_{q=1}=\frac{e^{\beta U}e^{-\beta
H_i}}{e^{\beta U}\tr e^{-\beta H_i}}= \frac{e^{-\beta H_i}}{\tr
e^{-\beta H_i}}. \ee

\nd It remains as yet to ascertain whether the entropy is indeed a
maximum, a point that has not been sufficiently discussed thus
far. Taking the second derivative of $L_T$ we easily obtain

\be \frac{\partial^2 L_T}{\partial p_i^2}=-q p_i^{q-2}, \ee i.e.,
for $q>0$ the entropy will be a maximum. {\it This result entails
that negative $q-$values are not acceptable}.

\subsection{General quantal  treatment of the Tsallis' 1988 version}
\nd Within the general quantum
 nonextensive framework, all information concerning the
 system of interest is provided by the statistical operator $\hat{\rho}$.
 To find it  one has to extremize the
 information
 measure \be
S_q=\frac{{\cal X}_q-1}{1-q}, \label{entropy} \ee where now
 \be
{\cal X}_q=\tr\,\left(\hat{\rho}^q\right),
 \ee
  subject
 to the
 normalization requirement
\be \tr\,(\hat{\rho}) =1 \ee and the assumed a priori knowledge
 of the expectation values (EV)s of, say,
 $M$ relevant observables $\hat{O}_j$ with $j=1...M$

 \be
\langle\hat{O}_j \rangle=\tr\,(\hat
 \rho \, \hat{O}_j).
 \ee
  The density or statistical operator $\hat \rho$ is here the quantum version of the
set of probabilities of the preceding Subsection. Following
exactly the same procedure as before, with $M$ Lagrange
multipliers $\lambda_j$, we obtain a density operator of the form

\be \hat{\rho}=\frac{\hat{f}^{1/(q-1)}}{\bar{Z}_q}, \label{rho}
\ee with a partition function given by \be
\bar{Z}_q=\tr\;\hat{f}^{1/(q-1)}, \label{Zq} \ee where \be
\hat{f}=1-(q-1)\sum_j \frac{\lambda_j}{q{\cal X}_q} \delta
\hat{O}_j \label{f} \ee is sometimes called  the configurational
characteristics. It is finite only  if its argument is positive.
$\hat{f}$ is null if its argument is negative (cut-off condition).
The quantities \be \delta
\hat{O}_j=\hat{O}_j-\langle\hat{O}_j\rangle_q \ee are usually
called centered operators.

\subsection{Connections between de 1988 and the TMP formalism}
\nd Comparing Eq. (\ref{rho}) with the density operator of TMP
treatment \cite{TMP}, we see that it is easy to pass from one to
other via an appropriate mapping. If the subindex $u$ stands for
the 1988 NET version we have

\ben \label{compqs} \left. q-1\right|_u &\mapsto&\left.
1-q\right|_{TMP} \cr \left. \lambda_j/q\right|_u&\mapsto&\left.
\lambda_j\right|_{TMP}. \een An interesting consequence of Eq.
(\ref{compqs}) ensues: as $S_q$ will be convex for $q|_u>0$ we
discover that the convex region for $S_q$ within  the TMP
formalism will be $q|_{TMP}<2$. Now, it has been shown in Ref.
\cite{OLM,Rlenzi} that in the TMP instance on also finds the
requirement $q>0$. Thus,  TMP and 1988 results should be valid for
$q\in(0,2)$.

\section{III. EQUIVALENCE OF THE DIFFERENT TSALLIS VERSIONS}

\nd As stated above  one takes, for a numerical treatment of the
canonical ensemble i) the internal energy $U_q$ as a numerical
datum and  ii) the energy spectrum. This provides one with a
probability distribution. We have called i)  and ii) the canonical
practitioner's input. We advance here an {\it alternative route}
(we call it the {\it parametric approach}), in which, {\it
instead} of giving $U_q$ as a numerical input data, one a priori
fixes the numerical value of an effective inverse temperature
$\beta^*$. With such an alternative procedure one is able to
achieve two different goals, {\it for a given energy-spectrum}:
\begin{itemize} \item
a unification of all Tsallis' techniques, from a theoretical
viewpoint, and \item  a very convenient numerical procedure, from
a practical perspective.
\end{itemize}

\nd Indeed, by recourse to an analysis similar to the one given
above, effected now for {\it all $4$ types} of Tsallis
distribution mentioned in the Introduction, it is easy to show
that this parametric technique can i) link the four NET versions
and ii) give the pertinent four distribution-forms a ``universal"
(common) appearance. The ensuing results are summarized in Table 1
for the canonical ensemble. The energy spectrum
$\{\varepsilon_i\}$ constitutes the basic a priori data. A
power-law distribution-form (common to all 4 Tsallis versions) of
exponent $q^*(q)$ can be built up, for any $q-$value, that reads

\be p_i= Z_q^{-1}\,\,\exp _{q^{\ast}(q)}(-\beta ^{\ast
}\varepsilon _{i});\,\,\,{\rm whose\,\,(version-dependent)\,\,
ingredients\,\,are:} \label{univers} \ee
\begin{itemize}\item an effective inverse temperature $\beta^*$,
given a priori as a number, \item  $Z_q,$ the normalization
constant (partition function), and
\item an ``effective" nonextensivity index $q^*=q^{*}(q)$, given in the
Table. Indeed, save for the case of the 1988 version, we have
$q^*=q$.
\end{itemize}
In all four cases, $U_q$ is evaluated a posteriori, as indicated
in the Table.  One can demonstrate that the appropriate  entropic
form $S_q$   is (a posteriori) maximized by (\ref{univers}) (for
{\it this} particular $U_q$). Also,
\begin{enumerate}
\item $T$ is the ``true" thermodynamic temperature that a
thermometer would measure, and
\item $\beta$ is the Lagrange multiplier of the putative MaxEnt
process one could have employed to reach the ``parametric" PD
(\ref{univers}) using the canonical practitioner's input.
\end{enumerate}  By TO we mean Tsallis' 1988 original treatment, and by
TMP-R the OLM one \cite{OLM,Rlenzi,FPS}.  {\it From the canonical
practitioner's viewpoint} $\beta^*$ is, save for the CT and OLM
(TMP-R) instances, a rather formidable quantity that  makes the
probability distribution a self-referential one (the signature of
the self-referential character is the appearance of
 $\mathcal{X}_q$ in the expression for $\beta^*$).
  {\it This is not so from the parametric perspective}. As explained
below, if one fixes $\beta^*$ at the outset, considering it as an
input parameter, the self-referential difficulty vanishes. One is
then able to numerically build in fast fashion the PD
(\ref{univers}) for very many $\beta^*-$values in the interval
$[0, \infty]$ and then ``read" off the Table the appropriate
values for $T (\beta),\,\,U_q,\,\,\mathcal{X}_q$ associated to any
given $\beta^*$.\vskip 3mm




\begin{tabular}{|c|c|c|c|c|}
\hline \multicolumn{5}{|c|}{Table 1: Non-extensive Statistics} \\
\hline \multicolumn{5}{|c|}{Probability Distribution Function} \\
\hline \multicolumn{5}{|c|}{$p_{i}=Z^{-1}\exp _{q^{\ast }}(-\beta
^{\ast }\varepsilon _{i}),$ if $\beta ^{\ast }\varepsilon
_{i}<\frac{1}{1-q^{\ast }}
$; \ $Z=\sum_{i}\exp _{q^{\ast }}(-\beta ^{\ast }\varepsilon _{i})$} \\
\hline \multicolumn{5}{|c|}{$p_{i}=0,$ \ \ \ \ \ \ \ \ \ \ \ \ \ \
\ \ \ \ \ \ \ \ \ \ \ \ if $\beta ^{\ast }\varepsilon
_{i}>\frac{1}{1-q^{\ast }}$}  \\ \hline & $\beta ^{\ast }$ &
$q^{\ast }$ & $U_{q}$ & $T$ \\ \hline $TO$ & $\beta ^{\ast
}=\frac{\beta }{q\chi _{q}+\left( q-1\right) \beta U_{q}^{TO}}$ &
$q^{\ast }=2-q$ & $U_{q}^{TO}=\sum_{i}p_{i}\varepsilon _{i}$
& $k_{B}T=\frac{k_{B}T^{\ast }-\left( q-1\right) U_{q}^{TO}}{q\chi _{q}}$ \\
\hline
$CT$ & $\beta ^{\ast }=\beta $ & $q^{\ast }=q$ & $U_{q}^{CT}=%
\sum_{i}p_{i}^{q}\varepsilon _{i}$ & $k_{B}T=k_{B}T^{\ast }$ \\ \hline
$TMP$ & $\beta ^{\ast }=\frac{\beta }{\chi _{q}+\left( 1-q\right) \beta
U_{q}^{TMP}}$ & $q^{\ast }=q$ & $U_{q}^{TMP}=\frac{\sum_{i}p_{i}^{q}%
\varepsilon _{i}}{\sum_{i}p_{i}^{q}}$ & $k_{B}T=\frac{k_{B}T^{\ast }-\left(
1-q\right) U_{q}^{TMP}}{\chi _{q}}$ \\ \hline
$TMP-R$ & $\beta ^{\ast }=\frac{\beta }{1+\left( 1-q\right) \beta
U_{q}^{TMP-R}}$ & $q^{\ast }=q$ & $U_{q}^{TMP-R}=\frac{\sum_{i}p_{i}^{q}%
\varepsilon _{i}}{\sum_{i}p_{i}^{q}}$ & $k_{B}T=k_{B}T^{\ast }-\left(
1-q\right) U_{q}^{TMP-R}$ \\ \hline
\end{tabular}

\begin{center}
\bigskip

\bigskip
\end{center}
\nd Notice  that, if  any one of the four Tsallis' probability
distributions (PDs) is available, we can with it, using the
appropriate equations of the version at hand, evaluate the
quantities $U_q$, $\mathcal{X}_q$, and from them obtain $T$. Now
for translating into another Tsallis' version we \newline
 -use our PD to evaluate the
$U_q-$value of the new version. With it and  \newline
 -the above values for $\mathcal{X}_q$ and
$T$, compute the {\it new} $\beta^*$, appropriate for this second
version.   \newline
 - Finally, employ Eq. (\ref{univers}).

\nd In order to gain some insight onto  the properties of each
Tsallis-version we have computed, for two different $q$ values,
namely, $q=0.6$ (Figs. 1 and 3) and $q=1.5$ (Figs. 2 and 4) the
relative internal energy $U_q/(h \nu)$ (Figs. 1 and 2) and the
specific heat $C_q/k_B$ (Figs. 3 and 4) for a system of equally
spaced levels labelled by a ``quantum number" $i$, like in the
case of the harmonic oscillator, for instance (inter-level
distance $=\varepsilon_0$),
with $i=0,1,2,...$ and $%
\varepsilon _{0}=h \nu. $ For all four Figs., the horizontal axis
represents $k_BT/(h \nu)$ and the $q=1$ results are also shown for
comparison. A glance to Figs. 1-4 shows that the four Tsallis
versions yield different results, and that the OLM instance is the
one that resembles the orthodox $q=1$ situation in closer fashion.
In particular, notice in Fig. 2 that the OLM$-q=1.5$ curve is
identical to that for $q=1$. Specific heat undulations are a
consequence of Tsallis' cut-off, as explained in Ref. \cite{FPS}.
The {\it asymptotic} behavior when T $\rightarrow \infty $ for
$U_{q}$ and $C_{q}$ is given by the following expressions:

$U_{q}\thicksim T^{\alpha }$

$C_{q}\thicksim T^{\alpha -1}$

\nd where the appropriate  $\alpha $-exponent values, for  each
Tsallis version, are shown below:

\bigskip

\begin{tabular}{|c|c|c|c|}
\hline $TO$ & $CT$ & $TMP$ & $TMP-R$ \\ \hline $\alpha
=\frac{1}{q}$ & $\alpha =2-q$ & $\alpha =\frac{1}{q}$ & $\alpha
=1$
\\ \hline
\end{tabular} \bigskip
\subsection{Advantages of the parametric form (\ref{univers})}

\nd It should be remarked that using the probability distribution
(PD) (\ref{univers}) exhibits the advantage of avoiding
self-reference no matter {\it which} of the normalization choices
one uses. In writing down the PD in the fashion

\be p_i= Z_q^{-1}\,\,\exp _{q^*}(-\beta^{\ast}\varepsilon
_{i});\,\,\,Z_q = \sum_i\,\exp _{q^{\ast}}(-\beta^{\ast
}\varepsilon _{i}), \label{univers1} \ee you can numerically
input, for a given  \ben q^* &=& 2-q;\,\,\,{\rm for
\,\,\,the\,\,\,1988\,\,\,version} \cr q^* &=& q; \,\,\,{\rm for
\,\,\,all\,\,\,the\,\,\,other\,\,\,versions}, \een  any
$\beta^*-$value whatsoever and you get the PD. No self-reference
here. Since the data are the spectrum, $\beta^*$, and $q^*$, we
speak of a ``parametric" calculation. With such a PD you evaluate
both $\mathcal{X}$ and $U_q$ and then proceed to ascertain the
value of the physical temperature $T$ using the relations given in
Table 1. From a numerical viewpoint therefore, (\ref{univers1})
provides one with a convenient tool to work out nonextensive
thermostatistics problems.

 \section{Conclusions} \nd We have revisited the four different
extant versions of Tsallis' nonextensive thermostatistics and
shown that they can be unified (Cf. Table 1) in the sense of
deriving the transformations that provide a link between any given
pair of versions. It must be emphasized, however, that  {\it
numerical results are different} for the distinct versions. In
particular, we have shown that the currently abandoned original
Tsallis version of 1988 is a quite legitimate one.

\vskip 3mm \nd {\bf FIGURE CAPTIONS} \vskip 4mm

{\bf Fig. 1:} Relative internal energy $U_q/h\nu$ vs. $k_BT/h\nu$
($q=0.6$) for the four different instances of
Tsallis-normalization (see text).

\vskip 3mm

{\bf Fig. 2:} Relative internal energy $U_q/h\nu$ vs. $k_BT/h\nu$
($q=1.5$) for the four different instances of
Tsallis-normalization (see text).

\vskip 3mm

{\bf Fig. 3:} Specific heat $C_q/k_B$ vs. $k_BT/h\nu$ ($q=0.6$)
for the four different instances of Tsallis-normalization (see
text).\vskip 3mm

{\bf Fig. 4:} Specific heat  $C_q/k_B$ vs. $k_BT/h\nu$
 ($q=1.5$) for the four different instances of Tsallis-normalization
(see text).

 
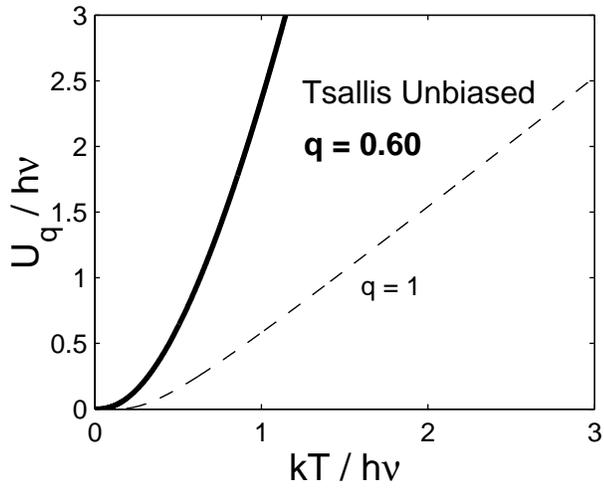
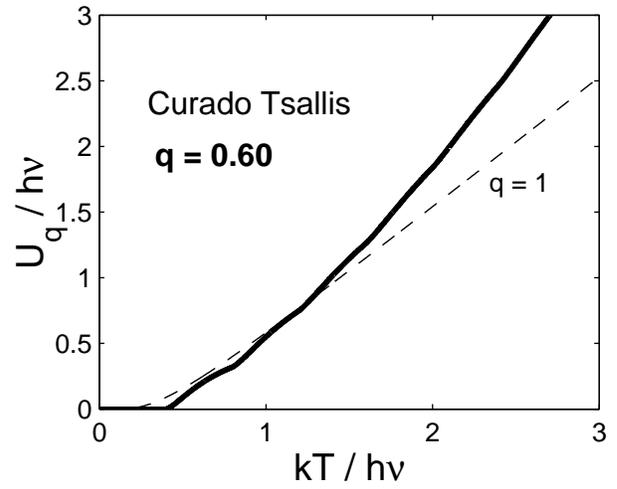
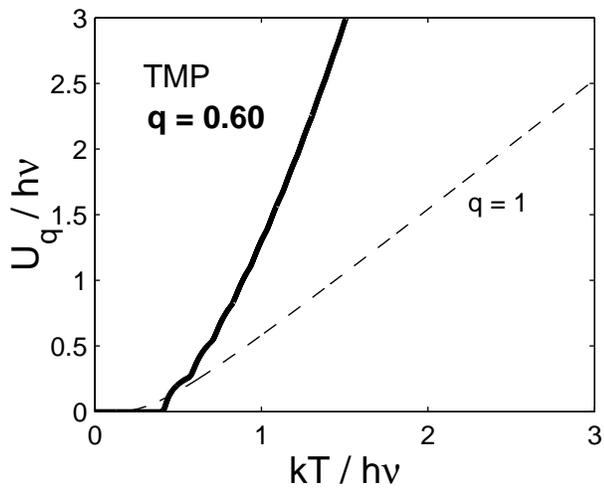
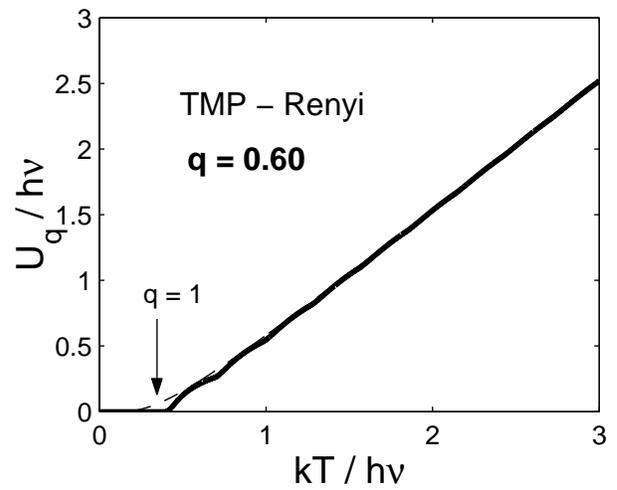
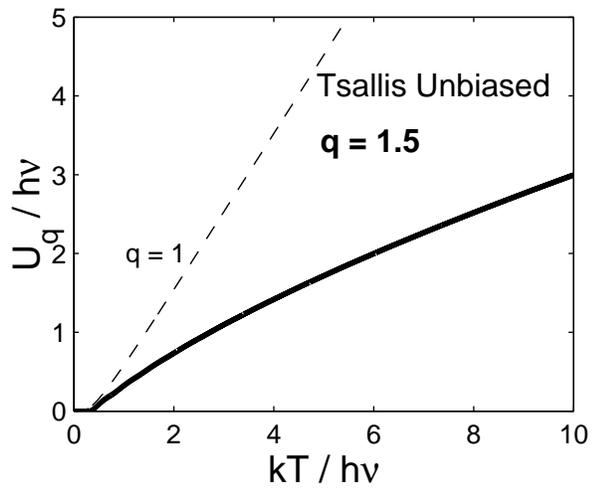
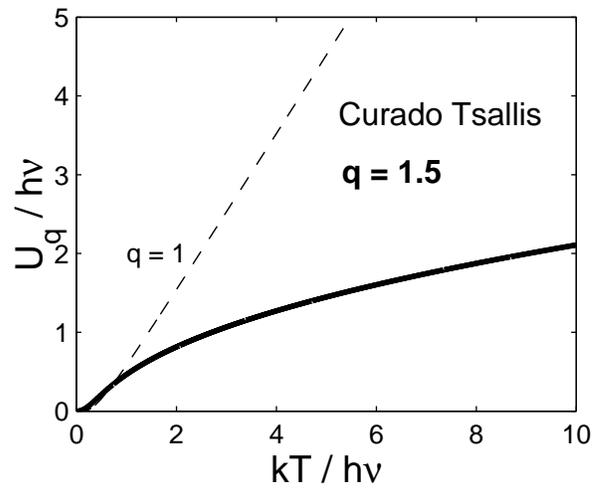
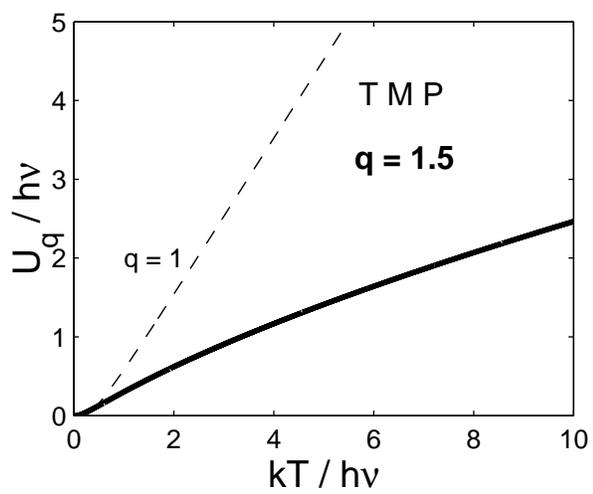
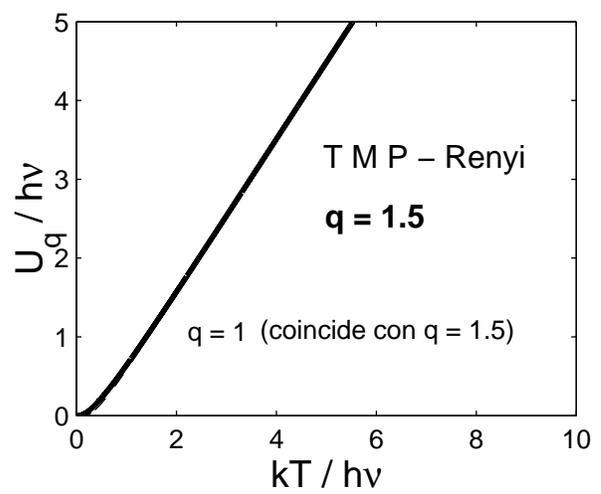
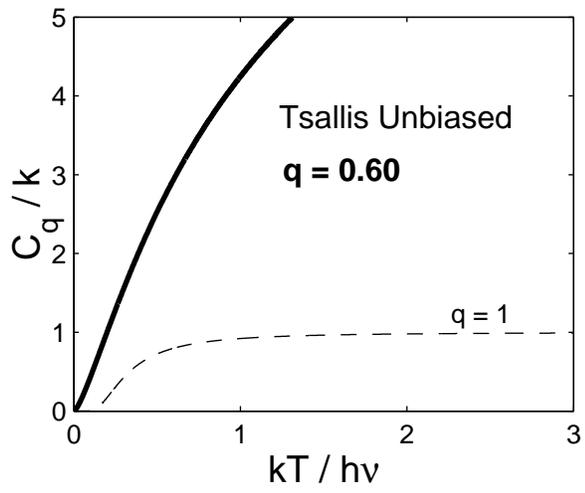
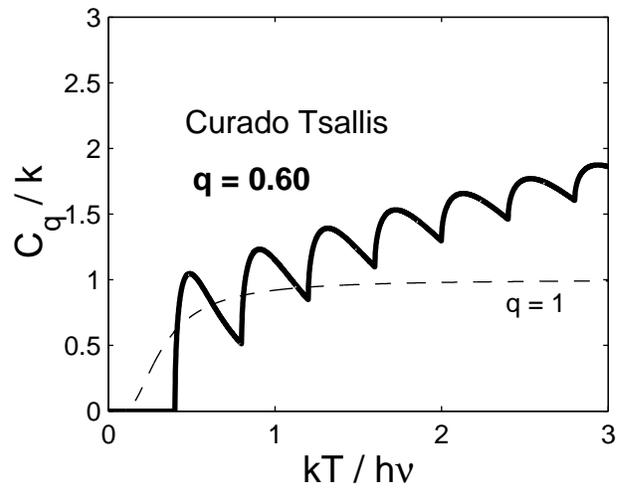
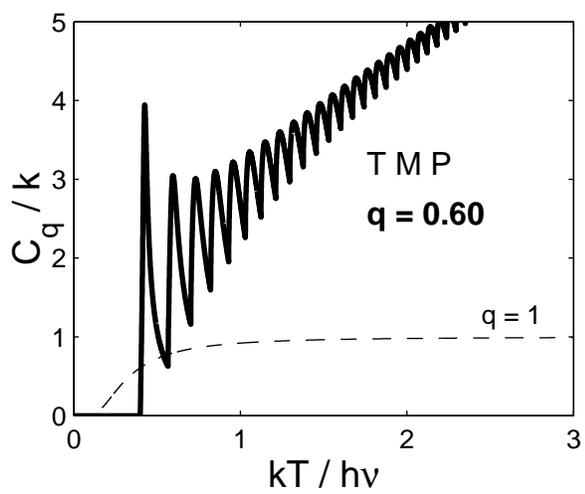
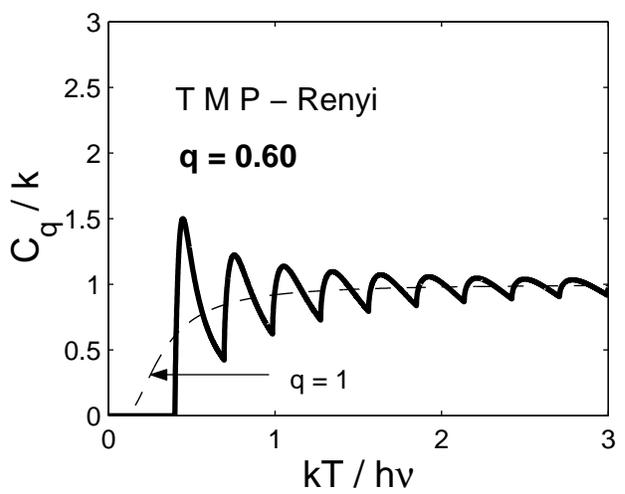
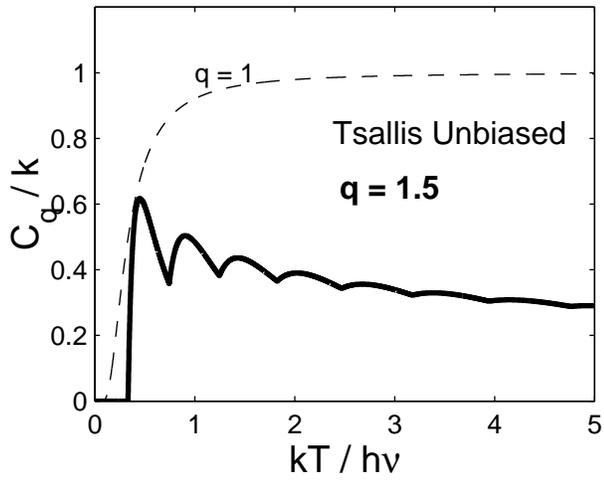
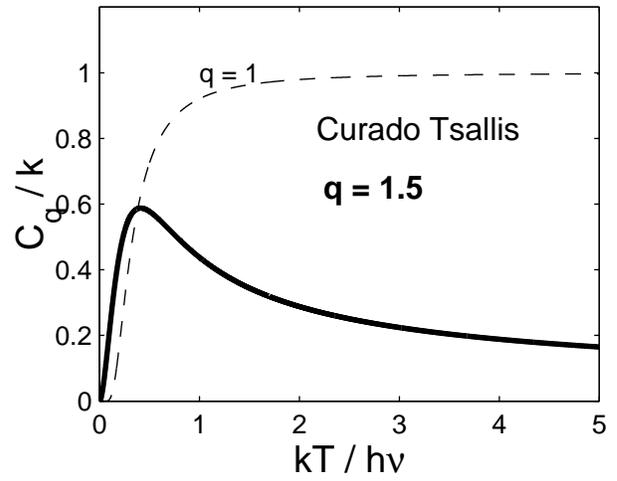
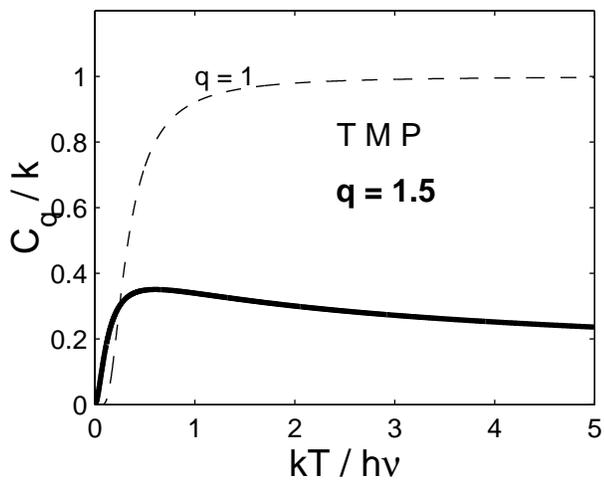
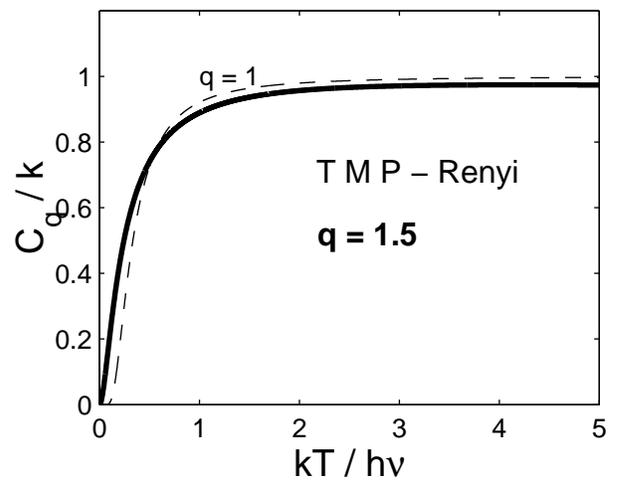
\end{document}